\documentclass[12pt]{article}
\usepackage{latexsym,equations}
\begin{document}
\renewcommand{\theequation}{\thesection.\arabic{equation}}
\def\prg#1{\medskip{\bf #1}}
\def\lra{\leftrightarrow}        \def\Ra{\Rightarrow}
\def\nin{\noindent}              \def\pd{\partial}
\def\dis{\displaystyle}          \def\dfrac{\dis\frac}
\def\grl{{GR$_\Lambda$}}         \def\vsm{\vspace{-10pt}}
\def\egt{{\scriptstyle\rm EGT}}  \def\Lra{{\Leftrightarrow}}
\def\ads3{AdS$_3$}               \def\cs{{\scriptscriptstyle \rm CS}}
\def\sm{{\scriptscriptstyle\rm M}}
\def\nb{{\Large $\bullet$} }     \def\bc{{\bar c}}
\def\Leff{\hbox{$\mit\L_{\hspace{.6pt}\rm eff}\,$}}
\def\bull{\raise.25ex\hbox{\vrule height.8ex width.8ex}}
\def\pr#1{{\tt #1}}

\def\cL{{\cal L}}     \def\cM{{\cal M }}    \def\cK{{\cal K}}
\def\cO{{\cal O}}     \def\cE{{\cal E}}     \def\cH{{\cal H}}
\def\tR{{\tilde R}}   \def\tG{{\tilde G}}   \def\hcO{\hat{\cal O}}
\def\hcH{\hat{\cH}}   \def\tI{{\tilde I}}   \def\hO{{\hat O}}
\def\bcH{{\bar\cH}}   \def\bcK{{\bar\cK}}   \def\bD{{\bar D}}
\def\bH{{\bar H}}     \def\D{{\Delta}}      \def\cC{{\cal C}}
\def\bL{{\bar L}}     \def\tIr{{\tilde\Ir}} \def\bI{{\bar I}}

\def\G{\Gamma}        \def\S{\Sigma}        \def\L{{\mit\Lambda}}
\def\a{\alpha}        \def\b{\beta}         \def\g{\gamma}
\def\d{\delta}        \def\m{\mu}           \def\n{\nu}
\def\th{\theta}       \def\k{\kappa}        \def\l{\lambda}
\def\vphi{\varphi}    \def\ve{\varepsilon}  \def\p{\pi}
\def\r{\rho}          \def\Om{\Omega}       \def\om{\omega}
\def\s{\sigma}        \def\t{\tau}          \def\tom{{\tilde\omega}}
\def\bw{{\bar w}}     \def\bC{{\bar C}}
\def\nn{\nonumber}
\def\be{\begin{equation}}             \def\ee{\end{equation}}
\def\ba#1{\begin{array}{#1}}          \def\ea{\end{array}}
\def\bea{\begin{eqnarray} }           \def\eea{\end{eqnarray} }
\def\beann{\begin{eqnarray*} }        \def\eeann{\end{eqnarray*} }
\def\beal{\begin{eqalign}}            \def\eeal{\end{eqalign}}
\def\lab#1{\label{eq:#1}}             \def\eq#1{(\ref{eq:#1})}
\def\bsubeq{\begin{subequations}}     \def\esubeq{\end{subequations}}
\def\bitem{\begin{itemize}}           \def\eitem{\end{itemize}}

\title{Black hole entropy
       from the boundary conformal structure in 3D gravity with torsion}

\author{M. Blagojevi\'c$^{1,2}$ and B. Cvetkovi\'c$^{1,}$\footnote{
        E-mail: mb@phy.bg.ac.yu and cbranislav@phy.bg.ac.yu} \\
\normalsize $^1$ Institute of Physics, P.O.Box 57,
                 11001 Belgrade, Serbia \\
\normalsize $^2$ Department of Physics, Univ. of Ljubljana,
                 Jadranska 19, 1000 Ljubljana, Slovenia}
\date{}
\maketitle

\begin{abstract}
Asymptotic symmetry of the Euclidean 3D gravity with torsion is
described by two independent Virasoro algebras with different central
charges. Elements of this boundary conformal structure are combined
with Cardy's formula to calculate the black hole entropy.
\end{abstract}


\section{Introduction} 
\setcounter{equation}{0}

Thermodynamic properties of black holes, the gravitational objects
with a highly complex dynamical structure, are expected to give us
important clues to the quantum nature of gravity \cite{1,2}. Of
particular importance for the development of these ideas have been
the outstanding contributions achieved in the context of
three-dimensional (3D) gravity \cite{3,4,5,6,7,8,9,10}, a simple
model for exploring basic features of the gravitational dynamics.
Traditionally, 3D gravity is formulated as general relativity with a
cosmological constant (\grl), with an underlying {\it Riemannian\/}
structure of spacetime. In the early 1990s, Mielke and Baekler
formulated a more general model---the model for 3D gravity with
torsion based on {\it Riemann-Cartan\/} geometry \cite{11}. Such an
approach might give us a new insight into the relationship between
geometry and the dynamical structure of gravity.

Further development along these lines led to a number of interesting
results \cite{12,13,14,15,16,x1,17,18,19}. In particular, it was shown
that: (a) the Mielke-Baekler model in Minkowskian spacetime possesses
the black hole solution, and (b) for suitable boundary conditions,
its asymptotic symmetry is described by two independent Virasoro
algebras with {\it different\/} central charges \cite{13,16,x1,18}.
One is now tempted to use this {\it Minkowskian\/} asymptotic structure
and calculate the black hole entropy in the manner of Strominger
\cite{10}. However, we prefer to base our study on the {\it Euclidean}
formalism, where Cardy's formula, needed in the calculation,
has a  natural statistical interpretation \cite{20,21}.

In the present paper, we study the Euclidean version of the
Mielke-Baekler (MB) model and show that its asymptotic symmetry is
locally {\it isomorphic\/} to the asymptotic symmetry of the Minkowskian
theory.  After this step, we are able to combine the resulting
asymptotic conformal structure with Cardy's formula for the
asymptotic density of states of a boundary CFT \cite{20,21} and
calculate the black hole entropy. The result is in complete agreement
with the calculations based on the gravitational partition function
\cite{19}. This agreement is well-known in \grl, but here, its validity
is extended to the general MB model.

The layout of the paper is as follows. In Sect. II, we present basic
aspects of the Euclidean MB model for 3D gravity with
torsion, including the form of the black hole solution. In Sect. III,
we discuss the asymptotic conditions and find the corresponding
restrictions on the original gauge parameters. In Sect. IV, we find
that the Poisson bracket algebra of the Euclidean theory is
asymptotically conformal, and in Sect. V, we combine the boundary
conformal structure with Cardy's formula to derive the black hole
entropy. Sect. VI is devoted to concluding remarks, while appendices
contain some technical details.

Our conventions are the same as in Ref. \cite{19}: the Latin indices
$(i,j,k,...)$ refer to the local orthonormal frame, the Greek indices
$(\m,\n,\r,...)$ refer to the coordinate frame, and both run over
$0,1,2$; $\eta_{ij}=(+,+,+)$ are metric components in the local
frame; totally antisymmetric tensor $\ve^{ijk}$ and the related
tensor density $\ve^{\m\n\r}$ are both normalized by $\ve^{012}=+1$.

\section{Euclidean 3D gravity with torsion} 
\setcounter{equation}{0}

We begin with a brief account of the MB model in the Euclidean formalism,
as a preparation for our treatment of the black hole entropy via Cardy's
formula \cite{19,20}.

Euclidean 3D gravity with torsion can be formulated as a gauge theory
of the Euclidean group $ISO(3)$. In this approach, gauge potentials
corresponding to local translations and local rotations are the triad
field $b^i$ and the spin connection $\om^i$ (1-forms), and the
corresponding field strengths are the torsion and the curvature
(2-forms): $T^i:=db^i+\ve^i{}_{jk}\om^j\wedge b^k$,
$R^i:=d\om^i+\frac{1}{2}\,\ve^i{}_{jk}\om^j\wedge\om^k$. In local
coordinates $x^\m$, the gauge potentials are represented as
$b^i=b^i{_\m}dx^\m$, $\om^i=\om^i{_\m}dx^\m$, and gauge
transformations are parametrized by $\xi^\m$ and $\th^i$:
\bea
\d_0 b^i{_\m}&=&
   -\ve^i{}_{jk}b^j{}_{\m}\th^k-(\pd_\m\xi^\r)b^i{_\r}
   -\xi^\r\pd_\r b^i{}_\m \, ,                             \nn\\
\d_0\om^i{_\m}&=&-\nabla_\m\th^i-(\pd_\m\xi^\r)\om^i{_\r}
   -\xi^\r\pd_\r\om^i{}_\m \, ,                            \lab{2.1}
\eea
where $\nabla_\m\th^i=\pd_\m\th^i+\ve^i{}_{jk}\om^j{_\m}\th^k$.
The metric has the form $g=\eta_{ij} b^i\otimes b^j$, with
$\eta_{ij}=$ diag$(1,1,1)$.

The geometric structure of $ISO(3)$ gauge theory corresponds to {\it
Riemann-Cartan\/} geo\-me\-try (for more details, see, for instance,
Refs. \cite{22,23}).

\prg{The action integral.} Mielke and Baekler proposed
a topological model for 3D gravity in Riemann-Cartan spacetime
\cite{11}, which is a natural generalization of \grl. The model is
based on the action
\bsubeq\lab{2.2}
\be
\bI=aI_1+\L I_2+\a_3I_3+\a_4I_4+I_m\, ,                    \lab{2.2a}
\ee
where $I_m$ is a matter contribution, and
\bea
&&I_1:= 2\int b^i\wedge R_i\, ,                             \nn\\
&&I_2:=-\frac{1}{3}\,\int\ve_{ijk}b^i\wedge b^j\wedge b^k\,,\nn\\
&&I_3:=\int\left(\om^i\wedge d\om_i
  +\frac{1}{3}\ve_{ijk}\om^i\wedge\om^j\wedge\om^k\right)\,,\nn\\
&&I_4:=\int b^i\wedge T_i\, .                               \lab{2.2b}
\eea
\esubeq
Here, the first two terms are of the same form as in \grl, $a=1/16\pi G$
and $\L$ is the cosmological constant, $I_3$ is the Chern-Simons action
for the connection, and $I_4$ is a torsion counterpart of $I_1$.

The {\it Euclidean\/} MB action \eq{2.2} is obtained from its {\it
Minkowskian\/} counterpart \cite{18} by the process of analytic
continuation, and consequently, the parameters $(a,\L,\a_3,\a_4)$ are
expressed in terms of the corresponding Minkowskian quantities, as
shown in Appendix A. The standard form of the action \eq{2.2} is
convenient for calculations, but at the end, we shall always use
\eq{A2} to return to the Minkowskian parameters.

In the sector $\a_3\a_4-a^2\ne 0$, the vacuum field equations are
non-degenerate:
\bsubeq\lab{2.3}
\be
2T^i=p\ve^i{}_{jk}\,b^j\wedge b^k\, ,\qquad
  2R^i=q\ve^i{}_{jk}\,b^j\wedge b^k\, ,                    \lab{2.3a}\\
\ee
where
\be
p:=\frac{\a_3\L+\a_4 a}{\a_3\a_4-a^2}\, ,\qquad
  q:=-\frac{(\a_4)^2+a\L}{\a_3\a_4-a^2}\, .                \lab{2.3b}
\ee
\esubeq

Introducing the Lavi-Civita connection $\tom^i$ by
$db^i+\ve^i{}_{mn}\tom^m b^n=0$, the field equations imply that
Riemannian piece of the curvature $R^i(\tom)$ has the form
\be
2R^i(\tom)=\Leff\,\ve^i{}_{jk}\,b^j\wedge b^k\, ,\qquad
\Leff:= q-\frac{1}{4}p^2\, ,                               \lab{2.4}
\ee
where $\Leff$ is the effective cosmological constant. Thus, our
spacetime has maximally symmetric metric; it is known as the
hyperbolic 3D space $H^3$, and has the isometry group $SO(3,1)$.
In what follows, we will restrict our attention to the sector with
positive $\Leff$, which represents the Euclidean continuation of
anti-de Sitter space \cite{19}. Accordingly, we use the notation
$\Leff=:1/\ell^2$.

\prg{The black hole solution.} For $\Leff>0$, equation \eq{2.4} has a
well known solution for the metric, the Euclidean BTZ black hole
\cite{6,8}. In Schwarzschild-like coordinates $x^\m=(t,r,\varphi)$, the
metric has the form
\bea
&&ds^2=N^2dt^2+N^{-2}dr^2+r^2(d\varphi+N_\vphi dt)^2\, ,    \nn\\
&&N^2:=\left(-8Gm+\frac{r^2}{\ell^2}-\frac{16G^2J^2}{r^2}\right)\, ,
  \qquad N_\vphi:=-\frac{4GJ}{r^2}.                         \lab{2.5}
\eea
The zeros of $N^2$, $r_+$ and $r_-=-i\r_-$, are related to the black
hole parameters $m$ and $J$ by the relations
$r^2_+-\r^2_-=8Gm\ell^2$, $r_+\r_-=4GJ\ell$, both $\vphi$ and $t$ are
periodic,
$$
0\le\vphi<2\pi\, ,\qquad 0\le t<\b\, ,\qquad
  \b=\frac{2\pi\ell^2 r_+}{r_+^2+\r_-^2}\, ,
$$
and the black hole manifold is topologically a solid torus.

Starting with the BTZ metric \eq{2.5}, one can find the pair ($b^i$,
$\om^i$) which solves the field equations \eq{2.3}, and represents
the Euclidean black hole with torsion \cite{19}:
\bsubeq\lab{2.6}
\bea
&&b^0= Ndt\, ,\qquad b^1=N^{-1}dr\, , \qquad
  b^2=r\left(d\vphi+ N_\vphi dt\right)\, ,                 \lab{2.6a}\\
&&\om^i=\tom^i+\frac{p}{2}\,b^i\, ,                        \lab{2.6b}
\eea
where the Levi-Civita connection $\tom^i$ reads:
\be
\tom^0=Nd\vphi\, ,\qquad \tom^1=-N^{-1}N_\vphi dr\, ,\qquad
  \tom^2=-\frac{r}{\ell^2}dt+ rN_\vphi d\vphi\, .          \lab{2.6c}
\ee
\esubeq

Formally, the substitution $8Gm=-1,J=0$, ``transforms'' the black hole
solution into the hyperbolic space $H^3$, the Euclidean analogue of AdS$_3$.
These solutions are locally isometric, but they have different topological
properties.

\section{Asymptotic conditions} 
\setcounter{equation}{0}

Asymptotic conditions are an essential element of every field theory.
In the canonical formalism, they are needed for the construction of
well-defined symmetry generators and the related conserved charges.

\prg{Asymptotic configuration.} Our choice of the asymptotic
conditions in the MB model is essentially based on the arguments
formulated twenty years ago by Brown and Henneaux in the context of
\grl\ \cite{4}: they should be (a) sufficiently general to include the
black hole configuration and (b) allow for the action of SO(3,1), the
isometry group of $H^3$, but (c) sufficiently regular to yield
well-defined canonical generators.

The asymptotic configuration of the triad field $b^i{_\m}$ that satisfies
(a) and (b) is given by
\bsubeq\lab{3.1}
\be
b^i{_\m}= \left( \ba{ccc}
   \dfrac{r}{\ell} +\cO_1 & \cO_4  & \cO_1  \\
   \cO_2 & \dfrac{\ell}{r}+\cO_3   & \cO_2  \\
   \cO_1 & \cO_4                   & r+\cO_1
                 \ea
          \right)   \, .                                   \lab{3.1a}
\ee
Here, for any $\cO_n=c/r^n$, we assume that $c$ is {\it not a
constant\/}, but a function of $t$ and $\vphi$, $c=c(t,\vphi)$, which
is the simplest way to ensure the global $SO(3,1)$ invariance. This
assumption is crucial for a highly non-trivial structure of the
resulting asymptotic symmetry.

The asymptotic form of $\om^i{_\m}$ is chosen in accordance with
the field equations:
\be
\om^i{_\m}=\left( \ba{ccc}
   \dfrac{pr}{2\ell}+\cO_1 & \cO_4  & \dfrac{r}{\ell}+\cO_1 \\
   \cO_2 & \dfrac{p\ell}{2r}+\cO_3  & \cO_2                 \\
   -\dfrac{r}{\ell^2}+\cO_1 & \cO_4 & \dfrac{pr}{2}+\cO_1
                  \ea
                  \right)  \, .                            \lab{3.1b}
\ee
\esubeq
A verification of the third condition (c) is left for the next section.

\prg{Asymptotic parameters.} Having chosen the asymptotic conditions,
we now wish to find the subset of gauge transformations \eq{2.1} that
respect these conditions. They are defined by restricting the
original gauge parameters in accordance with \eq{3.1}, which yields:
\bsubeq\lab{3.2}
\bea
&&\xi^0=\ell\left[ T
  -\frac{1}{2}\left(\frac{\pd^2 T}{\pd t^2}\right)
              \frac{\ell^4}{r^2}\right] +\cO_4\, ,         \nn\\
&&\xi^2=S-\frac{1}{2}\left(\frac{\pd^2 S}{\pd\vphi^2}\right)
              \frac{\ell^2}{r^2}+\cO_4\, ,                 \nn\\
&&\xi^1=-\ell\left(\frac{\pd T}{\pd t}\right)r+\cO_1\, ,   \lab{3.2a}
\eea
and
\bea
&&\th^0=\frac{\ell^2}{r}
        \left(\frac{\pd^2 T}{\pd t\pd\vphi}\right)+\cO_3\,,\nn\\
&&\th^2=-\frac{\ell^3}{r}
        \left(\frac{\pd^2 T}{\pd t^2}\right)+\cO_3\, ,     \nn\\
&&\th^1=\left(\frac{\pd T}{\pd\vphi}\right)+\cO_2\, .      \lab{3.2b}
\eea
\esubeq
Here, $T$ and $S$ are functions that satisfy the conditions
\bsubeq
\be
\ell\frac{\pd S}{\pd t}=-\frac{\pd T}{\pd\vphi}\, ,\qquad
\frac{\pd S}{\pd\vphi}=\ell\frac{\pd T}{\pd t}\, ,
\ee
which can be interpreted as the Cauchy-Riemann equations of complex
analysis. Indeed, after introducing the complex variables
$$
w^\pm:=\vphi\pm i\frac{t}{\ell}\, ,\qquad T^\pm:=S\pm iT\, ,
$$
one can rewrite these conditions as
\be
\pd_-T^+=0\, ,\qquad \pd_+ T^-=0\, .
\ee
\esubeq
Consequently, general solutions for $T^\pm$ have the form
$T^+=g(w^+)$, $T^-=h(w^-)$, where $g,h$ are some analytic/anti-analytic
functions.

\prg{The commutator algebra.} The commutator of two gauge
transformations \eq{2.1} is closed: $[\d_0',\d_0'']=\d_0'''$, where
$\d_0'=\d_0(\xi',\th')$ and so on. The composition of parameters
reads:
\bea
&&\xi'''{}^\m=\xi'{}^\r\pd_\r\xi''{}^\m
              -\xi''{}^\r\pd_\r\xi'{}^\m\, ,                \nn\\
&&\th'''{}^i=\ve^i{}_{mn}\th'{}^m\th''{}^n
             +\xi'\cdot\pd\th''{}^i-\xi''\cdot\pd\th'{}^i\,.\nn
\eea
Applying the composition law to the restricted parameters \eq{3.2} and
keeping only the lowest-order terms, one finds the relations
\be
T'''{}^\pm=T'{}^\pm\pd_2 T''{}^\pm-T''{}^\pm\pd_2T'{}^\pm\,.\lab{3.4}
\ee
If we introduce the notation
\bea
&&\d_n^{-}:=\d_0(T^-=e^{inw^-},T^+=0)\, ,                   \nn\\
&&\d_n^{+}:= -\d_0(T^+=e^{-in w^+},T^-=0)\, ,               \nn
\eea
the commutator algebra takes the form
\be
[\d_n^{\mp},\d_m^{\mp}]=-i(n-m)\d_{n+m}^{\mp}\, .           \lab{3.5}
\ee

In general, the commutator algebra implies that the commutator of two
$(T,S)$ transformations produces not only a $(T,S)$ transformation,
but also an additional, pure gauge transformation (for which
$T=S=0$). This result motivates us to introduce an improved
definition of the asymptotic symmetry: it is the symmetry defined by
the parameters \eq{3.2}, modulo the pure gauge transformations.
Locally, this symmetry coincides with the conformal symmetry on the
2-dimensional torus. The subalgebra generated by $n,m=0,\pm 1,$
is $so(3,1)$, as expected.

\section{Canonical realization of the asymptotic symmetry} 
\setcounter{equation}{0}

In this section, we use the canonical formalism to study the
Poisson bracket algebra of the asymptotic symmetry.

\subsection{The improved canonical generator}

Following the standard Hamiltonian analysis, one can construct the gauge
generator $G$, which produces the correct gauge transformations of
all phase-space variables \cite{24}. The transformation law of the
fields, defined by $\d_0\phi:=\{\phi\,,G\}$, is in complete
agreement with the gauge transformations \eq{2.1} {\it on shell\/}
\cite{19}.

The gauge generator $G$ is constructed as a spatial integral of
linear combinations of the first-class constraints, so that it
vanishes on the constraint hypersurface: $G\approx 0$. Moreover, it
acts on dynamical variables via the Poisson bracket operation, which
is defined in terms of functional derivatives. On a manifold with boundary,
$G$ does not have well-defined functional derivatives, but the problem
can be cured by adding suitable surface terms \cite{25}. The improved
canonical generator $\tG$ for the Euclidean MB model has the form:
\be
\tG=G+\G\, ,\qquad
\G:=-\int_0^{2\pi}d\vphi
         \left(\xi^0\cE^1+\xi^2\cM^1\right)\, ,            \lab{4.1}
\ee
where
\bea
&&\cE^1=
   2\left[\left(a+\frac{\a_3p}{2}\right)\om^0{}_2
  +\left(\a_4+\frac{ap}{2}\right)b^0{}_2-\frac{a}{\ell}b^2{}_2
  -\frac{\a_3}{\ell}\om^2{}_2\right]b^0{}_0\, ,            \nn\\
&&\cM^1=
  2\left[\left(a+\frac{\a_3p}{2}\right)\om^2{}_2
  +\left(\a_4+\frac{ap}{2}\right)b^2{}_2+\frac{a}{\ell}b^0{}_2
  +\frac{\a_3}{\ell}\om^0{}_2\right]b^2{}_2\, .            \nn
\eea
The adopted asymptotic conditions guarantee differentiability and
finiteness of $\tG$. Moreover, $\tG$ is also conserved.

\prg{Conserved charges.} The value of the improved generator $\tG$
defines the {\it gravitational charge\/}. Since $\tG\approx \G$, the
charge is completely determined by the boundary term $\G$. Note that
$\G=0$ for the pure gauge transformations.

For $\xi^2=0$, $\tG$ reduces to the time translation generator, while
for $\xi^0=0$ we obtain the spatial rotation generator. The
corresponding surface terms, calculated for $\xi^0=1$ and $\xi^2=1$,
respectively, have the meaning of {\it energy\/} and {\it angular
momentum\/}:
\be
E=-\int_0^{2\pi}d\vphi\,\cE^1 \, ,\qquad
M=-\int_0^{2\pi}d\vphi\,\cM^1 \, .                         \lab{4.2}
\ee
Energy and angular momentum are conserved gravitational charges.
Their values for the black hole configuration (3.2) are
\bsubeq\lab{4.3}
\be
E=i\left[m+\frac{\a_3}{a}\left(\frac{pm}{2}
          -\frac{J}{\ell^2}\right)\right]\, ,\qquad
M=i\left[J+\frac{\a_3}{a}
           \left(\frac{pJ}{2}+m\right)\right]\, .          \lab{4.3a}
\ee
The conserved charges in the MB model are linear combinations of the
\grl\ charges $m,J$. Returning to the Minkowskian parameters with the
help of Appendix A, we find
\be
E=i\left[m+\frac{\a_3}{a}\left(\frac{pm}{2}
          -\frac{J}{\ell^2}\right)\right]^\sm\, ,\qquad
M=-\left[J+\frac{\a_3}{a}
           \left(\frac{pJ}{2}-m\right)\right]^\sm\, ,      \lab{4.3b}
\ee
\esubeq
or equivalently, $E=iE^\sm$ and $M=-M^\sm$.

\subsection{Canonical algebra}

In the canonical formalism, the asymptotic symmetry is
determined by the Poisson bracket algebra of the improved generators
$\tG$. In the notation $G':=G[T',S']$, $G'':=G[T'',S'']$, and so on,
the Poisson bracket algebra is found to have the form
\bsubeq\lab{4.4}
\be
\left\{\tG'',\,\tG'\right\} =\tG''' + C''' \, ,            \lab{4.4a}
\ee
where the parameters $(T''', S''')$  are determined by the
composition rules  \eq{3.4}, and $C'''$ is the classical
{\it central term\/}:
\bea
C'''=&&~-(2a+\a_3p)\ell\int_0^{2\pi}d\vphi
    \left(\pd_2S''\pd_2^2T'-\pd_2S'\pd_2^2T''\right)       \nn\\
&&~+2\a_3\int_0^{2\pi}d\vphi
    \left(\pd_2T''\pd_2^2T'-\pd_2S''\pd_2^2S'\right)\, .   \lab{4.4b}
\eea
\esubeq

The Poisson bracket algebra \eq{4.4} can be brought into a more familiar
form by introducing the generators
\bea
&&L^-_n:=-\tG(T^-=e^{inw^-},T^+=0)\, ,                     \nn\\
&&L^+_n:=\tG(T^+=e^{-inw^+},T^-=0)\, ,                     \lab{4.5}
\eea
which produce the transformations with parameters $T^-=e^{inw^-}$
and $T^+=e^{-inw^+}$, respectively. These generators obey two independent
Virasoro algebras with central charges:
\bsubeq\lab{4.6}
\bea
\{L^\mp_m,L^\mp_n\}&=&-i(m-n)L^\mp_{n+m}
                      -\frac{ic^\mp}{12}m^3\d_{m+n,0}\, ,  \lab{4.6a}\\
ic^\mp&:=&24\pi\left[a\ell
     +\a_3\left(\frac{p\ell}2\mp i\right)\right]\, ,       \nn
\eea
and $\{L^-_n,L^+_m\}=0$. The algebra \eq{4.6a} is a central extension of
the commutator algebra \eq{3.5}.
Returning to the Minkowskian parameters by using \eq{A2}, we find
\be
c^\mp=24\pi\left[a\ell
      +\a_3\left(\frac{p\ell}{2}\mp 1\right)\right]^\sm\, ,
\ee
\esubeq
which are exactly the values found earlier in the Minkowskian formalism
\cite{18}. The boundary CFT is characterized by {\it different\/}
classical central charges, corresponding to holomorphic and
antiholomorphic sectors. The asymptotic Poisson bracket algebra for
the Euclidean MB theory is {\it isomorphic\/} to the corresponding
Minkowskian algebra.

\section{Black hole entropy via Cardy's formula} 
\setcounter{equation}{0}

Using the definitions \eq{4.5}, we can express the values of $L^\mp_0$
in terms of the conserved charges:
\be
L^\mp_0\approx \mp\frac{1}{2}(M\pm i\ell E)
        =\frac{1}{2}(\ell E^\sm\pm M^\sm)=:h^\mp\, .
\ee

Now, we can apply Cardy's formula for the asymptotic density of states of
a boundary CFT and calculate the black hole entropy. The formula is based
{\it entirely\/} on the asymptotic symmetry structure expressed by
the Virasoro algebra \eq{4.6}, and has the form:
\be
S=2\pi\sqrt{\frac{h^-{c^-}}6}+2\pi\sqrt{\frac{h^+{c^+}}6}\,,\lab{5.2}
\ee
where $h^\mp$ are given in terms of the Minkowskian parameters as
$$
h^\mp=(\ell m\pm J)^\sm\frac{c^\mp}{48\pi a\ell}
      =\frac{(r_+\pm r_-)^2}{8G\ell}\frac{c^\mp}{48\pi a\ell}\, .
$$
Thus, the entropy of the black hole with torsion takes the form
(in units $\hbar=1$):
\bea
S&=&\frac{\pi}{6\ell}\left[r_+(c^-+c^+)+r_-(c^--c^+)\right] \nn\\
 &=&\frac{2\pi r_+}{4G}+4\pi^2\a_3\left(pr_+-\frac{2r_-}\ell\right)\, .
                                                            \lab{5.3}
\eea
This result, obtained from the boundary conformal structure,
{\it coincides\/} with the gravitational entropy, based on the
calculation of the grand canonical partition function \cite{19}.

In \grl, where $\a_3=0$, we have $S={2\pi r_+}/{4G}$, as
expected. For $p=0$, our expression for $S$ reduces to Solodukhin's
result \cite{26}, obtained in Riemannian geometry, but with
a Chern-Simons term in the action.

Note that the second term in $S$, which represents the modification of
the \grl\ result, depends on both the outer and inner horizons, $r_+$
and $r_-$. Since the Euclidean black hole manifold does not contain
the inner horizon, its appearance in $S$ can be understood as a consequence
of the analytic structure of the black hole solution: the Euclidean
black hole ``remembers'' that its Minkowskian counterpart has the inner
horizon. On the other hand, the presence of the Chern-Simons coupling
constant $\a_3$ and the strength of torsion $p$ shows the influence of
the geometric structure of spacetime on the gravitational dynamics.

Cardy's formula \eq{5.2} holds under the following assumptions \cite{21}:
\bitem
\item[(a)] the boundary CFT is unitary: $h^-\geq 0,h^+\geq 0$
(this implies $c^-,c^+\ge 0$);
\vsm\item[(b)] the values $h^-$, $h^+$ are sufficiently
large: $h^-\gg c^-/24$, $h^+\gg c^+/24$;
\vsm\item[(c)] the lowest possible values of $h^-,h^+$ are zero:
$h^-_0=0,h^+_0=0$.
\eitem
With $c_0=24\pi a\ell=3\ell/2G$, the conditions (b) can be rewritten
in the form $\ell m\pm J \gg {c_0}/{12}$,
which is the same as in \grl. Consequently, in the semiclassical regime
$m$ has to be large in Planck units:
$$
m\gg 1/\hbar G\, .
$$

Clearly, the fact that $S$ and $c^\mp$ are not allowed to take negative
values imposes certain bounds on $\a_3,p$ and $r_\pm$.

Cardy's formula \eq{5.2} is obtained from the zero-loop (classical)
approximation of the CFT partition function. The one-loop correction
is obtained in a simmilar manner \cite{27}; when applied to the MB model,
it gives the one-loop correction to $S$ (Appendix B).

\section{Concluding remarks} 

In this paper, we found that the asymptotic symmetry of the Euclidean
3D gravity with torsion is the conformal symmetry, described by two
independent Virasoro algebras with different central charges and
locally isomorphic to the corresponding Minkowskian structure. The
black hole entropy is then calculated using  Cardy's formula for the
asymptotic density of states of a boundary CFT. The result is in
perfect agreement with the gravitational entropy, obtained via the
gravitational partition function \cite{19}. This agreement,
well-known for Riemannian \grl, is now generalized to the MB model of
3D gravity with torsion.

\section*{Acknowledgements} 

This work was supported by the Serbian and Slovenian Science Foundations.

\appendix
\section{Euclidean continuation} 
\setcounter{equation}{0}

In the process of analytic continuation, different terms of the
Minkowskian MB action $I_M$ transform into their Euclidean counterparts
according to the rule (Appendix A in \cite{19}):
\bea
&&b^iR_i\mapsto ib^iR_i\, ,\qquad
  b^0b^1b^2\mapsto -ib^0b^1b^2\, ,                           \nn\\
&&\cL_\cs(\om)\to\cL_\cs(\om)\, ,\qquad b^iT_i\to -b^iT_i\, .\lab{A1}
\eea
Consequently, the Euclidean action \eq{2.2} is obtained from $I_M$
through the analytic continuation, $I_M\mapsto\bI$, if its parameters
$(a,\L,\a_3,\a_4)$ are related to the Minkowskian parameters
$(a^\sm,\L^\sm,\a_3^\sm,\a_4^\sm)$ as follows:
\be
\ba{ll}
    a=ia^\sm\, ,   \\
    \a_3=\a_3^\sm\, ,
\ea \qquad
\ba{l}
   \L=-i\L^\sm\, , \\
   \a_4=-\a_4^\sm\, .
\ea                                                         \lab{A2}
\ee
We also have $J^\sm=-iJ$.
Note that \eq{A2} differs from the result defined by the analytic
continuation $I_M\mapsto iI_E$, discussed in Ref. \cite{19}, by
the presence of an additional factor $i$.

\section{The one-loop correction} 
\setcounter{equation}{0}

The one-loop correction to the black hole entropy can be found using
the results of Ref. \cite{27}, according to which the corrected Cardy
formula for the density of states has the form
$$
\r(h^-,h^+)=\r(h^-)\r(h^+)\, ,\qquad
\r(h^\mp)\approx\left[\frac{C^\mp}{96(h^\mp)^3}\right]^{1/4}
   \exp\left(2\pi\sqrt{\frac{h^\mp C^\mp}{6}}\right)\, ,
$$
where $C^\mp$ is the full central charge, $C^\mp=c^\mp$ + quantum
corrections. In the approximation $C^\mp\approx c^\mp$, taking the
logarithm of $\r(h^-,h^+)$ leads to
\be
S=\frac{1}{\hbar}S^{(0)}
  -\frac{3}{2}\ln\left(\frac{2\pi r_+}{\hbar G}\,{\ell\k}\right)
  +{\rm const}+\cO(\hbar)\, ,                              \lab{B1}
\ee
where $S^{(0)}$ is the classical value \eq{5.3}, and
$\k=(r_+^2-r_-^2)/\ell^2 r_+$ is is the surface gravity.
The first-order correction in \eq{B1} has the same form as in \grl.

\end{document}